\documentclass[conference]{IEEEtran}

\usepackage{graphicx}
\usepackage{balance}

\usepackage{algorithm}
\usepackage{algpseudocode}
\usepackage{subfig}

\let\labelindent\relax
\usepackage{enumitem}
\usepackage{multirow}

\begin{document}

\title{Container-Based Cloud Virtual Machine Benchmarking}

\author{\IEEEauthorblockN{Blesson Varghese, Lawan Thamsuhang Subba, Long Thai and Adam Barker}
\IEEEauthorblockA{School of Computer Science, University of St Andrews, UK, KY16 9SX\\
Email: \{varghese, lts4, ltt2, adam.barker\}@st-andrews.ac.uk}}

\author{
    \IEEEauthorblockN{Blesson Varghese\IEEEauthorrefmark{1}, Lawan Thamsuhang Subba\IEEEauthorrefmark{2}, Long Thai\IEEEauthorrefmark{2}, Adam Barker\IEEEauthorrefmark{2}}
    \IEEEauthorblockA{\IEEEauthorrefmark{1}School of Electronics, Electrical Engineering and Computer Science, Queen's University Belfast, UK
    \\varghese@qub.ac.uk}
    \IEEEauthorblockA{\IEEEauthorrefmark{2}School of Computer Science, University of St Andrews, UK
    \\\{lts4, ltt2, adam.barker\}@st-andrews.ac.uk}
}

\maketitle

 \begin{abstract}
With the availability of a wide range of cloud Virtual Machines (VMs) it is difficult to determine which VMs can maximise the performance of an application. Benchmarking is commonly used to this end for capturing the performance of VMs. Most cloud benchmarking techniques are typically \textit{heavyweight} - time consuming processes which have to benchmark the entire VM in order to obtain accurate benchmark data. Such benchmarks cannot be used in real-time on the cloud and incur extra costs even before an application is deployed. 

In this paper, we present \textit{lightweight} cloud benchmarking techniques that execute quickly and can be used in near real-time on the cloud. The exploration of lightweight benchmarking techniques are facilitated by the development of DocLite - Docker Container-based Lightweight Benchmarking. DocLite is built on the Docker container technology which allows a user-defined portion (such as memory size and the number of CPU cores) of the VM to be benchmarked. DocLite operates in two modes, in the first mode, containers are used to benchmark a small portion of the VM to generate performance ranks. In the second mode, historic benchmark data is used along with the first mode as a hybrid to generate VM ranks. The generated ranks are evaluated against three scientific high-performance computing applications. The proposed techniques are up to 91 times faster than a heavyweight technique which benchmarks the entire VM. It is observed that the first mode can generate ranks with over 90\% and 86\% accuracy for sequential and parallel execution of an application. The hybrid mode improves the correlation slightly but the first mode is sufficient for benchmarking cloud VMs.
\end{abstract}
\begin{IEEEkeywords}
cloud benchmarking; Docker; containers; lightweight benchmark; hybrid benchmark
\end{IEEEkeywords}

\IEEEpeerreviewmaketitle

\section{Introduction}
\label{introduction}
The cloud computing marketplace has become crowded with competitors each offering a wide-range of Virtual Machines (VMs) varying in their performance. With numerous options available it is challenging for a user to select VMs which maximise the performance of an application on the cloud. This results in applications under performing and increasing running costs on the cloud. 

Benchmarking is a technique that is commonly adopted to tackle the above problem in which the performance attributes of cloud VMs are captured \cite{cloudbenchmark-1, cloudbenchmark-2}. Benchmark data is then used to help rank the performance of VMs \cite{cloudbenchmark-3, cloudbenchmark-4}. However, cloud benchmarking methods are typically \textit{heavyweight} - by this we mean time consuming processes to benchmark an entire VM, which incur significant monetary costs. Heavyweight benchmarks \cite{cloudbenchmark-3} are usually obtained by running the benchmarking tool on the entire VM. For example, the cr1.8xlarge VM with 244GiB RAM will require over 10 hours to be benchmarked. This time consuming process results in costs even before an application is deployed on the VM. This definition of \textit{heavyweight} will be used throughout the paper. 

In this paper, we explore \textit{lightweight} cloud benchmarking techniques - processes which execute quickly and can be used in near real-time to collect metrics from cloud providers and VMs. In order to facilitate our exploration of lightweight benchmarking techniques we introduce DocLite - Docker Container-based Lightweight Benchmarking. DocLite is built around the Docker\footnote{http://www.docker.com/} container technology \cite{ cont-0a, cont-0, cont-1, cont-3}, which allows a user-defined portion (such as memory size and the number of CPU cores) of the VM to be benchmarked. For example, containers can be used to benchmark 1GB of a VM that has 256GB RAM. This has the core advantage that cloud VMs can be rapidly benchmarked for use in real-time, which in turn helps to reduce benchmarking costs for the purposes of comparison. Two important research questions that arise are: (i) how fast can our lightweight technique execute compared to heavyweight techniques which benchmark the entire VM? and (ii) how accurate will the generated benchmarks be? 

DocLite organises the benchmark data into four groups, namely memory and process, local communication, computation and storage. A user of DocLite provides as input a set of four weights (ranging from 0 to 5), which indicate how important each of the groups are to the application that needs to be deployed on the cloud. The weights are mapped onto the four benchmark groups and are used to generate a score for ranking the VMs according to performance. DocLite has two modes of operation. In the first mode containers are used to benchmark a small portion of a VM to generate the performance ranks of VMs. In the second mode, historic benchmark data are used along with the first mode as a hybrid in order to generate VM ranks. 

Three scientific case study applications are used to validate the benchmarking techniques. The experiments highlight that the lightweight technique (a) is up to 91 times faster than the heavyweight technique, and (b) generates rankings with over 90\% and 86\% accuracy for sequential and parallel execution of applications when compared to time consuming heavyweight techniques, which benchmark the whole VM. A small improvement is obtained in the quality of rankings for the hybrid technique. The key observation is that lightweight techniques are sufficient for benchmarking cloud VMs. 


This paper makes the following research contributions:
\begin{enumerate}
\item[i.] the development of lightweight cloud benchmarking techniques that can benchmark VMs in near real-time for generating performance ranks.
\item[ii.] the development of DocLite, a tool that incorporates the lightweight cloud benchmarking techniques.
\item[iii.] an evaluation using containers of varying sizes of VM memory against the whole VM.
\item[iv.] an evaluation of the accuracy of the benchmarks generated by DocLite on three scientific case study applications.
\end{enumerate}

The remainder of this paper is organised as follows. 
Section \ref{relatedwork} considers related research.
Section \ref{benchmarking} proposes two cloud benchmarking techniques.
Section \ref{implementation} considers a tool that is developed to incorporate the benchmarking techniques.  
Section \ref{studies} is an experimental evaluation of the proposed techniques using three case study applications. 
Section \ref{conclusions} concludes this paper by reporting future work.  

\section{Related Work}
\label{relatedwork}
Benchmarking is used to quantify the performance of a computing system \cite{rel-1, rel-1a}. Standard benchmarks such as Linpack are used for ranking the top supercomputers \cite{rel-2}. However, there are no standard benchmarking techniques accepted by the cloud community and there are numerous ongoing efforts to develop such benchmarks \cite{rel-3, rel-4, rel-5}. Benchmarking offers insight into a number of resources and services offered by the cloud \cite{rel-11a}. There is research exploring benchmarks for cloud databases \cite{cloudbenchmark-2}, for understanding the reliability and variability of cloud services \cite{rel-11, rel-11b}, for gathering network performance between cloud resources in workﬂows and web services \cite{rel-12, rel-13}. In this paper, benchmarking is performed directly on a VM. Consequently, the entire VM must be benchmarked for generating accurate benchmarks that can take a few hours to complete on large VMs. Currently, benchmarking methods that generate accurate and detailed benchmarks are heavyweight.

In order to facilitate cloud performance benchmarking in a meaningful way, benchmarking techniques need to be employed in near real-time and at the same time produce accurate benchmarks. This is because VMs have different performance characteristics over time and sometimes even in small time periods; Netflix uses a real-time benchmarking method for selecting VMs that match application requirements\footnote{http://www.brendangregg.com/blog/2015-03-03/performance-tuning-linux-instances-on-ec2.html}. Alternate virtualisation technology, such as containers with low boot up times and a high degree of resource isolation are likely to be the way forward to achieve lightweight techniques \cite{cont-2, cont-3}.

Preliminary experimental results already indicate that containers have lower overheads when compared to existing virtualisation technologies both for high-performance computing systems \cite{rel-6} as well as for the cloud \cite{rel-7}. Containers on the cloud is still in its infancy and recent research has reported employing containers for distributed storage \cite{rel-8}, reproducibility of research \cite{rel-9}, and in the context of security \cite{rel-10}. 

In this paper, we present novel techniques that employ container technology to benchmark cloud VMs in near real-time to produce reliable benchmarks. There is a large body of research on benchmarking, both at the resource and service levels. However, most benchmarking techniques at the resource level are limited in that they are time consuming and expensive, thereby restricting their use for actual deployments. The benefits of containers, such as resource isolation, are leveraged in the techniques we propose to achieve near real-time benchmarking of the VM on the cloud. 

\section{Container-Based Benchmarking}
\label{benchmarking}
In this section, two cloud benchmarking methods are proposed and presented. The first is a lightweight method that employs Docker container technology to benchmark cloud VMs in real-time. The second method, combines the use of benchmarks from the first along with historic benchmarks generated from heavyweight methods (benchmarking the entire VM) as a hybrid. The aim of both methods is to generate a ranking of cloud VMs based on performance. The benefit of using containers is that the amount of VM resources benchmarked, such as memory size and the number of CPU cores, can be limited. This benefit is leveraged such that only a small amount of resources available to a VM are benchmarked in comparison to heavyweight methods that benchmark the entire resource of a VM. 


A user can opt for either the lightweight or hybrid methods. A set of user defined weights, $W$ (considered later in this section), and historic benchmark data, $HB$, obtained from either a heavyweight method or from previous executions of the lightweight method can be provided as input. The lightweight benchmarks are obtained from \texttt{Obtain-Benchmark}, considered in Algorithm \ref{algorithm2}, which is represented as $B$. \texttt{Lightweight-Method} (Algorithm \ref{algorithm3}) takes as input $W$ and $B$ and \texttt{Hybrid-Method} (Algorithm \ref{algorithm4}) takes $HB$ as additional input.


Algorithm \ref{algorithm2} gathers the benchmarks from different cloud VMs. Consider there are $i = 1, 2, \cdots , m$ different VM types, and a VM of type $i$ is represented as $vm_{i}$. A container $c_{i}$ is created on each VM. In the context of cloud benchmarking, containers are used to facilitate benchmarking on different types of VMs by restricting the amount of resources used.

\begin{algorithm} 
	\caption{Obtain cloud benchmarks}
	\label{algorithm2}
	\begin{algorithmic}[1]
		\Procedure{Obtain\textendash Benchmark}{$mem$, $CPU\_cores$}{}
			\For{each virtual machine $vm_{i} \in VM$ }
				\State Create container $c_{i}$ of $mem$ size and $CPU\_cores$ on $vm_{i}$
				\State Execute standard benchmark tool on $c_{i}$
				\State Store benchmarks as $B$
		\EndFor
	\EndProcedure
	\end{algorithmic}
\end{algorithm}

Standard benchmark tools are executed using the container $c_{i}$. The latency and bandwidth information for a wide range of memory and process, computation, local communication and file related operations are collected. The benchmarks obtained for each $vm_{i}$ are stored in a file, $B$, for use by Algorithm \ref{algorithm3} and Algorithm \ref{algorithm4}. 

\subsection{Lightweight Method}
\label{lightweight}
The benchmarks need to be used to generate ranks of VMs based on performance. In the lightweight method, no historic benchmark data is employed. The benchmark data obtained from using containers, $B$, is used in this method as shown in Algorithm \ref{algorithm3}. 

Consider there are $j = 1, 2, \cdots , n$ attributes of a VM that are benchmarked, and $r_{i,j}$ is the value associated with each $j^{th}$ attribute on the $i^{th}$ VM. The attributes can be grouped as $G_{i, k} = \{r_{i, 1}, r_{i, 2}, \cdots \}$, where $i = 1, 2, \cdots m$, $k = 1, 2, \cdots , p$, and $p$ is the number of attribute groups. In this paper, four attribute groups are considered:

\begin{enumerate}
\item \textit{Memory and Process Group}, denoted as $G_{1}$ captures the performance and latencies of the processor. 
\item \textit{Local Communication Group} in which the bandwidth of both memory communications and interprocess communications are captured under the local communication group, denoted as $G_{2}$.
\item \textit{Computation Group}, denoted as $G_{3}$ which captures the performance of integer, float and double operations such as addition, multiplication and division and modulus. 
\item \textit{Storage Group} in which the file I/O related attributes are grouped together and denoted as $G_{4}$.
\end{enumerate}

\begin{algorithm} 
	\caption{Cloud ranking using lightweight method}
	\label{algorithm3}
	\begin{algorithmic}[1]
		\Procedure{Lightweight\textendash Method}{$W$, $B$}{}
			\State From $B$, organise benchmarks into groups, $G$
			\State Normalise groups, $\bar{G}$
			\State Score $VM$ using $\bar{G}.W$
			\State Generate performance ranks $R_p$ 
		\EndProcedure
	\end{algorithmic}
\end{algorithm}

The attributes of each group are normalised as $\bar{r}_{i, j} = \frac{r_{i, j} - \mu_{j}}{\sigma_{j}}$, where $\mu_j$ is the mean value of attribute $r_{i, j}$ over $m$ VMs and $\sigma_j$ is the standard deviation of the attribute $r_{i, j}$ over $m$ VMs. The normalised groups are denoted as $\bar{G}_{i, k} = \{\bar{r}_{i, 1}, \bar{r}_{i, 2}, \cdots \}$, where $i = 1, 2, \cdots m$, $k = 1, 2, \cdots , p$, and $p$ is the number of groups.

One input to Algorithm \ref{algorithm3} is $W$, which is a set of weights that correspond to each group (for the four groups $G_{1}, G_{2}, G_{3}, G_{4}$, the weights are $W=\{W_{1}, W_{2}, W_{3}, W_{4}\}$). For a given application, a few groups may be more important than the others. For example, if there are a large number of file read and write operations in a simulation, the Storage group represented as $G_{4}$ is important. The weights are provided by the user based on domain expertise and the understanding of the importance of each group to the application. Each weight, $W_{k}$, where $k=1, 2, 3, 4$ takes value between 0 and 5, where 0 indicates that the group is not relevant to the application and 5 indicates the importance of the group for the application. 

Each VM is scored as $S_{i} = \bar{G}_{i, k}.W_{k}$. The scores are ordered in a descending order for generating $Rp_{i}$ which is the ranking of the VMs based solely on performance. 

\subsection{Hybrid Method}
\label{hybrid}

The hybrid method employs benchmarks obtained in real-time, $B$, and historic benchmarks, $HB$, obtained either from a heavyweight method or a previous execution of the lightweight method as shown in Algorithm \ref{algorithm4}. This method accounts for past and current performance of a VM for generating ranks.

Attribute grouping and normalising are similar to those followed in Algorithm \ref{algorithm3}. The four groups used in the lightweight method are used here and the attributes are normalised using the mean and standard deviation values of each attribute.

When historic benchmarks are used, the method for grouping the attributes and normalising the groups are similar to what was followed previously. The attributes from historic benchmark data, $hr$, can be grouped as $HG_{i, k} = \{hr_{i, 1}, hr_{i, 2}, \cdots \}$, where $i = 1, 2, \cdots m$ for $m$ VM types, $k = 1, 2, \cdots , p$, and $p$ is the number of groups. Four groups, $HG_{1}$, $HG_{2}$, $HG_{3}$ and $HG_{4}$, are obtained. 

\begin{algorithm}[H]
	\caption{Cloud ranking using hybrid method}
	\label{algorithm4}
	\begin{algorithmic}[1]
		\Procedure{Hybrid\textendash Method}{$W$, $B$, $HB$}{}
			\State From $B$, organise benchmarks into groups, $G$
			\State Normalise groups, $\bar{G}$
			\State From $HB$, organise historic benchmarks into groups, $HG$
			\State Normalise groups, $\bar{HG}$
			\State Score $VM$ using $\bar{G}.W + \bar{HG}.W$
			\State Generate performance ranks $R_p$ 
		\EndProcedure
	\end{algorithmic}
\end{algorithm}

The attributes of each group are normalised as $\bar{hr}_{i, j} = \frac{hr_{i, j} - h\mu_{j}}{h\sigma_{j}}$, where $h\mu_j$ is the mean value of attribute $hr_{i, j}$ over $m$ VMs and $h\sigma_j$ is the standard deviation of the attribute $hr_{i, j}$ over $m$ VMs. The normalised groups are denoted as $\bar{HG}_{i, k} = \{\bar{hr}_{i, 1}, \bar{hr}_{i, 2}, \cdots \}$, where $i = 1, 2, \cdots m$, $k = 1, 2, \cdots , p$, and $p$ is the number of groups.

The set of weights supplied to the hybrid method are same as the lightweight method. Based on the weights each VM is scored as $S_{i} = \bar{G}_{i, k}.W_{k} + \bar{HG}_{i, k}.W_{k}$. The scores take the most current and historic benchmarks into account and are ordered in descending order for generating $Rp_{i}$, which is the performance ranking of the VMs.

The benchmarking methods will be affected if there are multiple workloads running on the same VM. Nonetheless, given a set of VMs which execute the benchmarking methods, the output of the methods will enable us to identify a VM or a subset of VMs that can meet the requirements of an application provided as a set of weights. The benchmarking methods considered in this section are incorporated into a tool presented in the next section. The methods are further evaluated against real world applications.

\section{DocLite Implementation}
\label{implementation}
In this section, the cloud platform and the VMs employed, the architecture of a tool that incorporates the benchmarking methods proposed in the previous section, and a sample of the benchmarks obtained on the cloud using the tool is presented. 

\subsection{Platform}
\label{implementation:cloudvms}

The Amazon Web Services (AWS) Elastic Compute Cloud (EC2)\footnote{http://aws.amazon.com/ec2/previous-generation/} is used to evaluate the benchmarking methods. The previous generation VMs (refer Table \ref{table1}) which have varying performance and become popular in the scientific community due to their longevity are chosen. 

\begin{table}[ht]
	\caption{Amazon EC2 VMs employed for benchmarking}
	\label{table1}
\begin{center}
	\begin{tabular}{c c p{0.7cm} p{2.3cm} p{0.9cm}}
		\hline	
		\textbf{VM Type}	&	\textbf{vCPUs}	&	\textbf{Memory (GiB)}	&	\textbf{Processor}	& \textbf{Clock (GHz)}	\\
		\hline	
		\hline	
		\texttt{m1.xlarge}	&	4	&	15.0	&	Intel Xeon E5-2650	&	2.00\\
		\texttt{m2.xlarge}	&	2	&	17.1	&	Intel Xeon E5-2665	&	2.40\\
		\texttt{m2.2xlarge}	&	4	&	34.2	&	Intel Xeon E5-2665	&	2.40\\
		\texttt{m2.4xlarge}	&	8	&	68.4	&	Intel Xeon E5-2665	&	2.40\\
		\texttt{m3.xlarge}	&	4	&	15.0	&	Intel Xeon E5-2670	&	2.60\\
		\texttt{m3.2xlarge}	&	8	&	30.0	&	Intel Xeon E5-2670	&	2.60\\
		\texttt{cr1.8xlarge}	&	32	&	244.0	&	Intel Xeon E5-2670	&	2.60\\		
		\texttt{cc2.8xlarge}	&	32	&	60.5	&	Intel Xeon X5570	&	2.93\\
		\texttt{hi1.4xlarge}	&	16	&	60.5	&	Intel Xeon E5620	&	2.40\\
		\texttt{hs1.8xlarge}&	16	&	117.0	&	Intel Xeon E5-2650	&	2.00\\
		\hline
	\end{tabular}
	\end{center}
\end{table}

The Docker container technology is used on the VMs for benchmarking. Docker is a portable and lightweight tool that facilitates the execution of distributed applications. It is advantageous in that container images require less storage space and consequentially deployment of containers is quick. Another useful feature of containers is resource isolation - the resources of a VM can be restricted to a specified amount of memory or number of cores (virtual CPUs) for benchmarking. The experiments were performed on 100MB, 500MB and 1000MB of RAM and on a single and on all vCPUs of the VM. In our approach Docker containers are used on top of the VMs and the resulting overhead is negligible as reported by industry experts\footnote{https://blogs.vmware.com/performance/2014/10/docker-containers-performance-vmware-vsphere.html}\footnote{https://blogs.vmware.com/performance/2015/05/running-transactional-workloads-using-docker-containers-vsphere-6-0.html}.

Container are an abstraction over the VM but at the same time as considered previously offer resource isolation. Process switching times captured using a container will effectively be that of the VM because no other container has access to the same VM resource. In other words, the workload executing on a container cannot interfere with the workload of another container executing on the same VM. 

One limitation of resource isolation in Docker containers is that I/O performance of the VM cannot be varied. Such a feature could have benefited cloud VM benchmarking when I/O bandwidth and latency measurements are accounted for. 

In this paper, the benchmarking tool employed is \texttt{lmbench} \cite{lmbench-1}. 
This tool was selected since (i) it is a single tool and can be easily deployed on the cloud, (ii) it provides a wide variety of benchmarks related to memory and process, computation, local communication and file related operations that capture the performance characteristics of the VM, and (iii) it has been employed for modelling the performance of cloud VMs reported in literature \cite{cloudbenchmark-1, cloudbenchmark-3, cloudbenchmark-100}. 

Two types of virtualisation are supported by Amazon VMs, namely \textit{paravirtual} and \textit{hvm}. Two AMIs corresponding to the virtualisation types were employed - \texttt{ami-9207e3a1} and \texttt{ami-b5120885}. The Docker image was created using the stock Ubuntu images from the Docker server, followed by the installation of lmbench. The image was uploaded to the Docker Hub repository and is available for public use\footnote{https://hub.docker.com/u/lawansubba/}.  

\subsection{Architecture}
\label{implementation:architecture}
\begin{figure}
	\centering
	\includegraphics[width=0.49\textwidth]{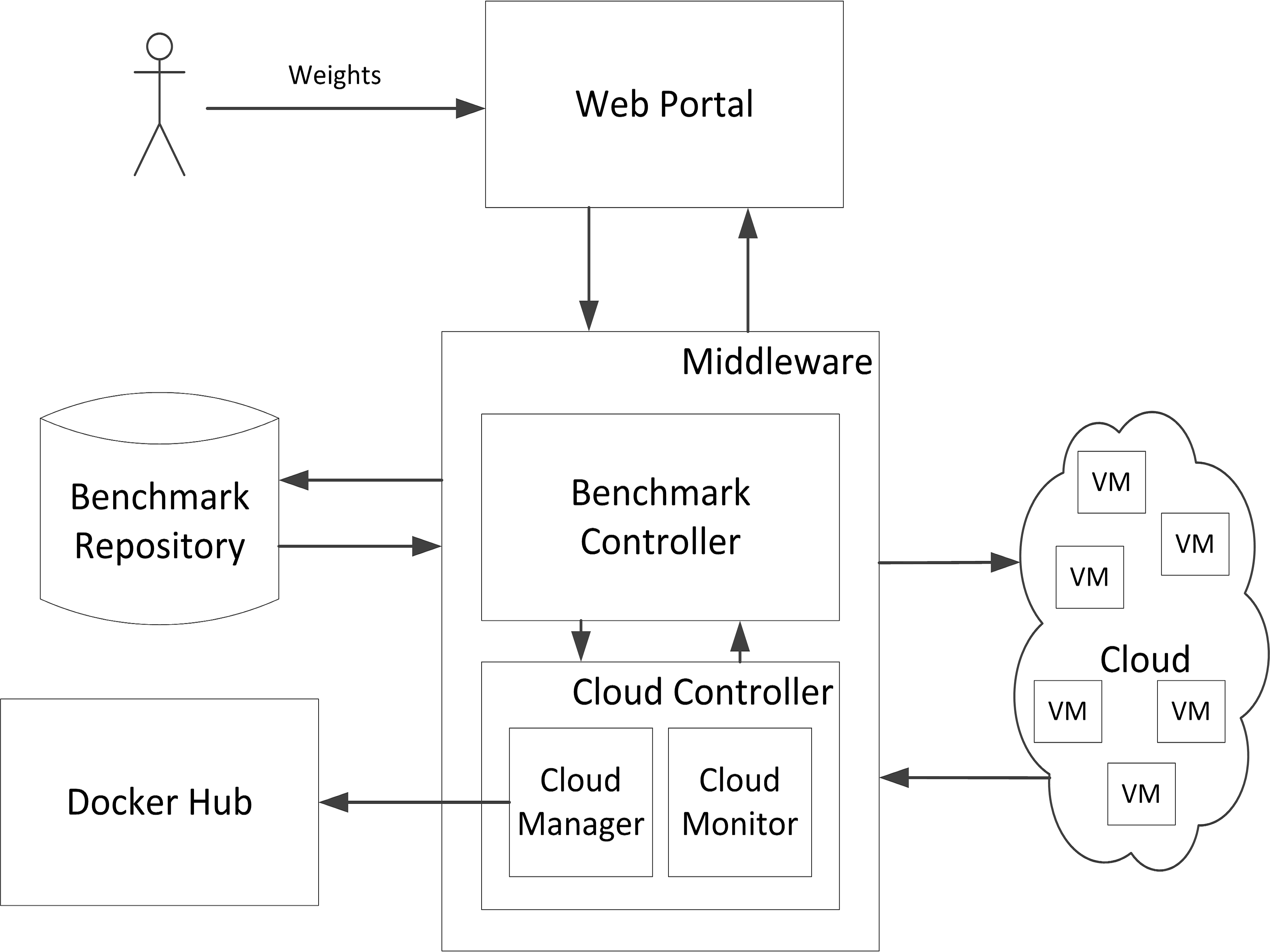}
	\caption{Architecture of DocLite}
	\label{fig:architecture}
\end{figure}

The container based benchmarking methods were implemented as a tool, we refer to as Docker Container-Based Lightweight Benchmarking tool (DocLite), which is available from https://github.com/lawansubba/DoCLite. The tool has three components, namely a web portal, a middleware and a benchmark repository as shown in Figure \ref{fig:architecture}.

\subsubsection{Web Portal}
The web portal is the user facing component developed using MVC.NET\footnote{http://www.asp.net/mvc} 
and Bootstrap\footnote{http://getbootstrap.com/}. A user provides a set of four weights $W=\{W_1, W_2, W_3, W_4\}$ that characterises the application to be deployed on the cloud as input; the amount of memory and number of cores to be benchmarked along with preferences of whether the benchmark needs to be executed sequentially or in parallel. The portal is also responsible for displaying the status of the cloud VMs that are used and the ranks generated from the benchmarks. In this paper, lmbench is used, however, the tool is flexible to accommodate other benchmarking tools and execute them independently or in any preferred combination.  

\subsubsection{Middleware}
The DocLite middleware comprises a Benchmark Controller and a Cloud Controller. The Benchmark Controller (i) incorporates the algorithms for lightweight and hybrid benchmarking considered in Section \ref{benchmarking}, (ii) pulls benchmark data from the repository for grouping and normalising the data, and (iii) generates the score for each VM based on the weights provided by the user. 

The Cloud Controller comprises of a Cloud Manager and a Cloud Monitor. The manager initiates cloud VMs and maintains them by executing the appropriate scripts for installing necessary packages and setting up Docker on the VM. The Docker images that are used are retrieved from the Docker Hub\footnote{https://hub.docker.com/} by the manager. The benchmarked data is deposited by the manager into the repository.

The monitor keeps track of the benchmarks that have started on the cloud VMs and reports the status of the VM to the portal. Monitoring is important to the Benchmark Controller to retrieve data from the repository after benchmarking.

\subsubsection{Repository}
The benchmark data obtained from the cloud is stored in a repository used by the Benchmark Controller for generating scores. Both historic and current benchmark data are stored in the repository. If the lightweight method is chosen, then the current benchmark data is used, where as if the hybrid method is chosen, then the historic data is used along with the current data. 

\subsection{Sample Benchmarks}
\label{implementation:benchmark}
\begin{figure*}[ht]
\centering
	\subfloat[Main memory latency]{\label{figure3a}\includegraphics[width=0.325\textwidth]{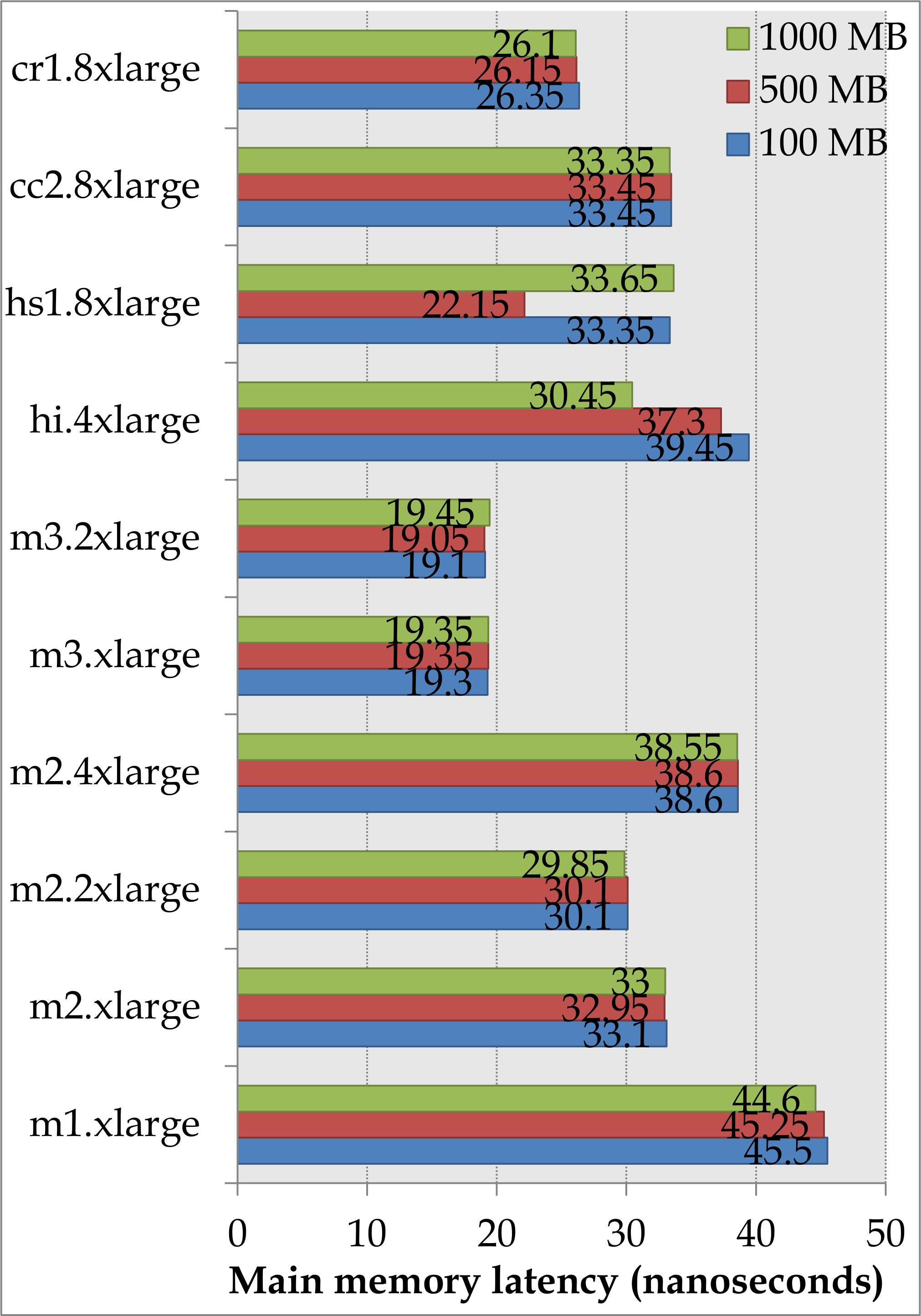}} \hfill
	\subfloat[Float division operation latency]{\label{figure3b}\includegraphics[width=0.325\textwidth]{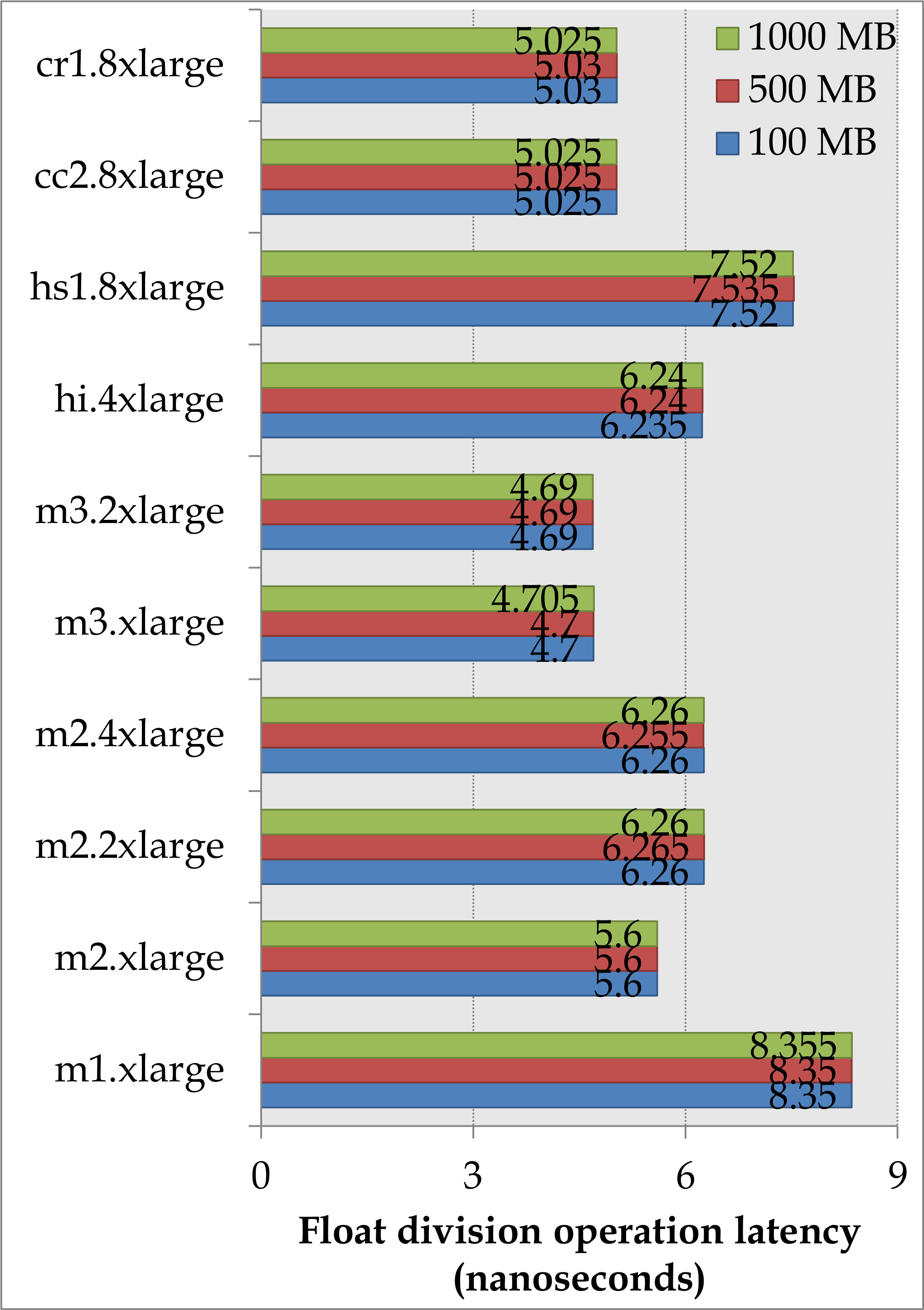}} \hfill
	\subfloat[Memory read bandwidth]{\label{figure3c}\includegraphics[width=0.325\textwidth]{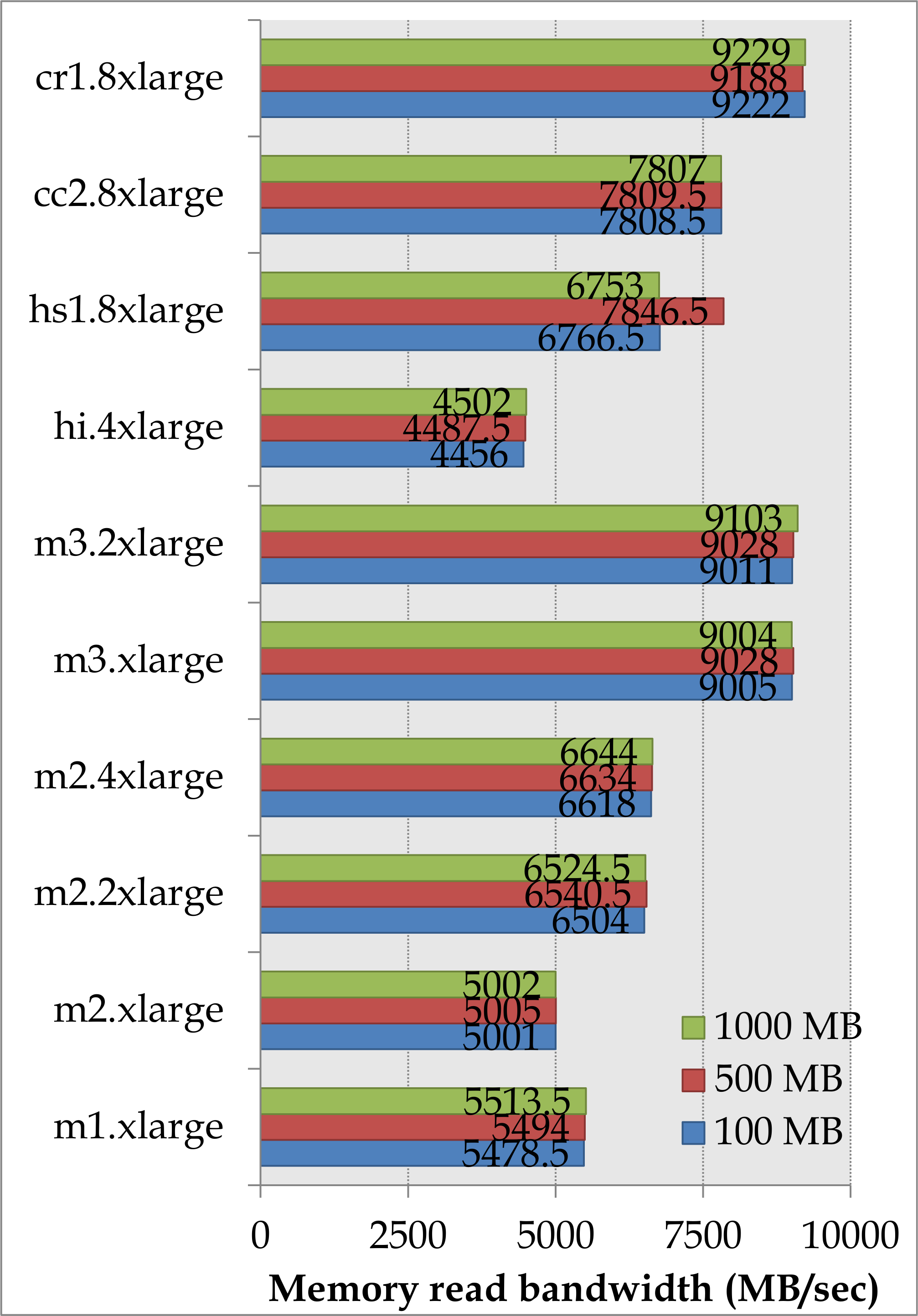}}
\caption{Sample lmbench benchmarks obtained from DocLite for 100MB, 500MB and 1000MB containers}
\label{figure3}
\end{figure*}

The VMs shown in Table \ref{table1} were benchmarked using DocLite by executing lmbench. Benchmarks for over fifty attributes related to memory and process, local communication, computation, and storage were obtained using containers of 100MB, 500MB and 1000MB. It is not within the scope of this paper to present all benchmarks. Therefore, a sample of three benchmarks is presented as shown in Figure \ref{figure3}.

Figure \ref{figure3a} shows the main memory latency for all VMs. It is immediately inferred that with the exception of \texttt{hs1.8xlarge} and \texttt{hi1.4xlarge} the main memory latencies are comparable for different container sizes. The exceptions are artefacts of measurements over networks. The best main memory performance is for the \texttt{m3} instances. Figure \ref{figure3b} shows the latency for a float division operation on each VM. Again, similar results are obtained for different container sizes. The bandwidth of memory read operations on all VMs are shown in Figure \ref{figure3c}. Maximum memory bandwidth is available on \texttt{cr1.8xlarge}. 

The key observation from the above three samples (also observed in all benchmarked data) is that there is a minimal difference on average of less than 2\% between the data points when a container of 100MB, 500MB or 1000MB is used. Given this small difference for different container sizes and the time taken to benchmark a VM using a small container is lower than a larger container, we hypothesise that (i) the benchmarking methods incorporated in DocLite can be employed in real-time, and (ii) VM rankings generated using the lightweight benchmarking methods will be comparable to heavyweight methods that benchmark the entire VM. This hypothesis will be evaluated in the next section using three case study applications. 

\section{Experimental Studies}
\label{studies}
In this section, three scientific case study applications are used to evaluate the benchmarking methods. An evaluation to validate the hypothesis of this research is considered by comparing the time taken to execute the benchmark on the VMs and comparing VM rankings generated by an empirical analysis, a heavyweight method and the two benchmarking methods. 

\begin{figure*}[ht]
	\centering
	\includegraphics[width=\textwidth]{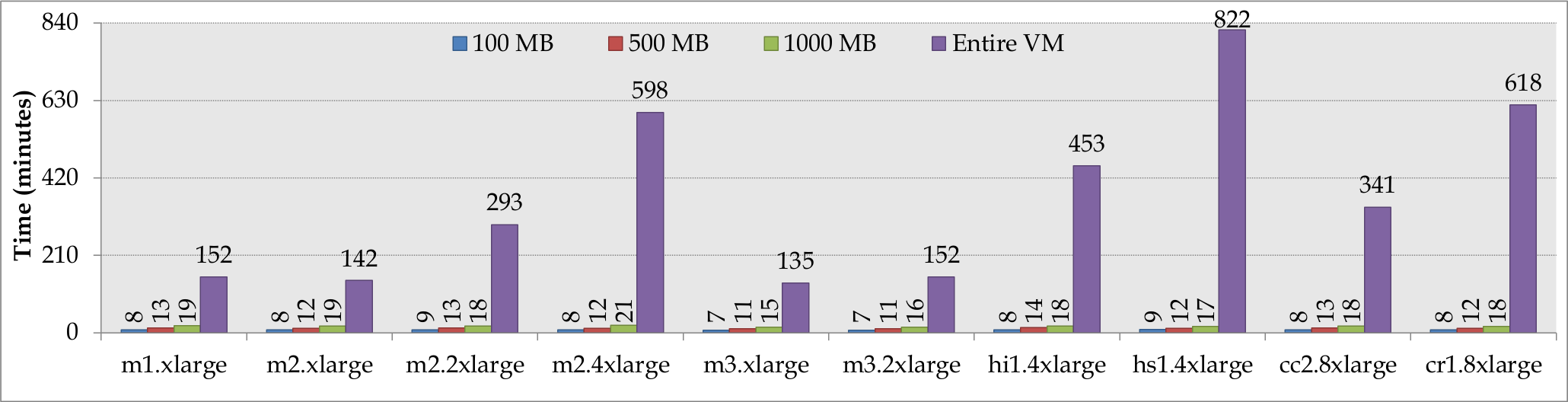}
	\caption{Time taken for executing the benchmarks using 100MB, 500MB and 1000MB containers and on the whole VM}
	\label{evaluation:feasibility}
\end{figure*}

\subsection{Case Study Applications}
\label{studies:casestudy}
Three high-performance computing applications are chosen to evaluate the benchmarking methods. These applications are executed on VMs as shown in Table \ref{table1} with at least 15 GiB memory so that the applications have sufficient memory on the VM. 

The first case study is a molecular dynamics simulation of a system comprising 10,000 particles in a three dimensional space used by theoretical physicists \cite{md-1}. The simulation solves differential equations to model particles for different time steps. The simulation is memory intensive with numerous read and write operations and computationally intensive requiring a large number of float operations. Local communication between processes are less relevant and the application does not require file operations. 

The second case study is a risk simulation that generates probable maximum losses due to catastrophic events \cite{risk-1}. The simulation considers over a million alternate views of a given year and a number of financial terms to estimate losses. The simulation is memory intensive with numerous read and write operations and at the same time computationally intensive requiring a large number of float operations to be performed both to compute the risk metrics. The local communication between processes are less relevant and the application does not require file operations.

The third case study is a block triagonal solver, which is a NASA Parallel Benchmark (NPB), version 3.3.1
\footnote{https://www.nas.nasa.gov/publications/npb.html}
\cite{npb-1}. This mathematical solver is used on a grid size of $162 \times 162 \times 162$ for 200 iterations. The solver is numerically intensive and memory and processor related operations are relevant, but does not take precedence over computations. Local communications and file operations have little effect on the solver. 

\subsection{Evaluation}
\label{studies:evaluation}

The aims of the experimental evaluation are to address two important research questions related to lightweight benchmarking. They are: 1) how fast can lightweight benchmarking execute compared to a heavyweight technique that benchmarks the entire VM? and 2) how accurate will the generated lightweight benchmarks be?

\subsubsection{Execution Time of Benchmarks}
\label{eval:time}
The first question related to speed is addressed by demonstrating the feasibility of the proposed lightweight benchmarking methods in real-time on the cloud. For this, the time taken to execute the lightweight and heavyweight benchmarking techniques are compared as shown in Figure \ref{evaluation:feasibility}. On an average the 100 MB, 500 MB, and 1000 MB containers take 8 minutes, 13 minutes and 18 minutes to complete benchmarking on all the VMs. Benchmarking the whole VM takes up to 822 minutes for \texttt{hs1.4xlarge}.  
It is immediately evident that container-based benchmarking is between 19-91 times faster than the benchmarking the entire VM.  

\subsubsection{Accuracy of Benchmarks}
\label{evaluation:empiricalanalysis}
The second question related to accuracy is addressed by evaluating the lightweight methods against three real-world case study applications. For this, ranks obtained from DocLite are compared against actual ranks of VMs when the application is executed. The following steps are used to evaluate the accuracy of the benchmarks:
\begin{itemize}
\item \textit{Step 1} - Execute case study application on all VMs.
\item \textit{Step 2} - Generate empirical ranks for the case study.
\item \textit{Step 3} - Provide weights of the application to DocLite.
\item \textit{Step 4} - Obtain benchmark ranks for the application.
\item \textit{Step 5} - Find correlation of benchmark and empirical ranks.
\end{itemize}

\begin{figure*}[ht]
\centering
	\subfloat[Case study 1 - sequential]{\label{figure1-1}\includegraphics[width=0.325\textwidth]{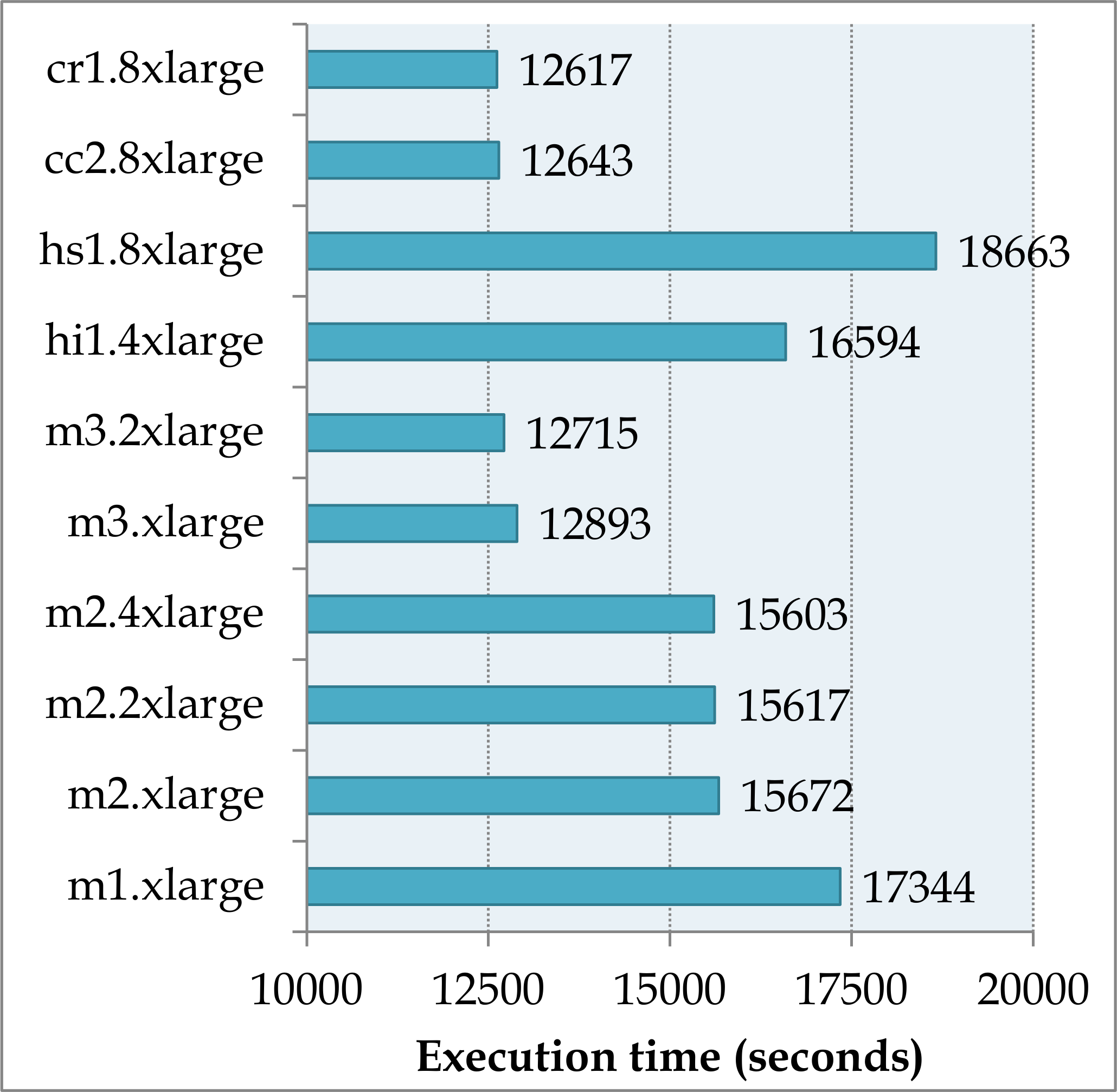}} \hfill
	\subfloat[Case study 2 - sequential]{\label{figure1-2}\includegraphics[width=0.325\textwidth]{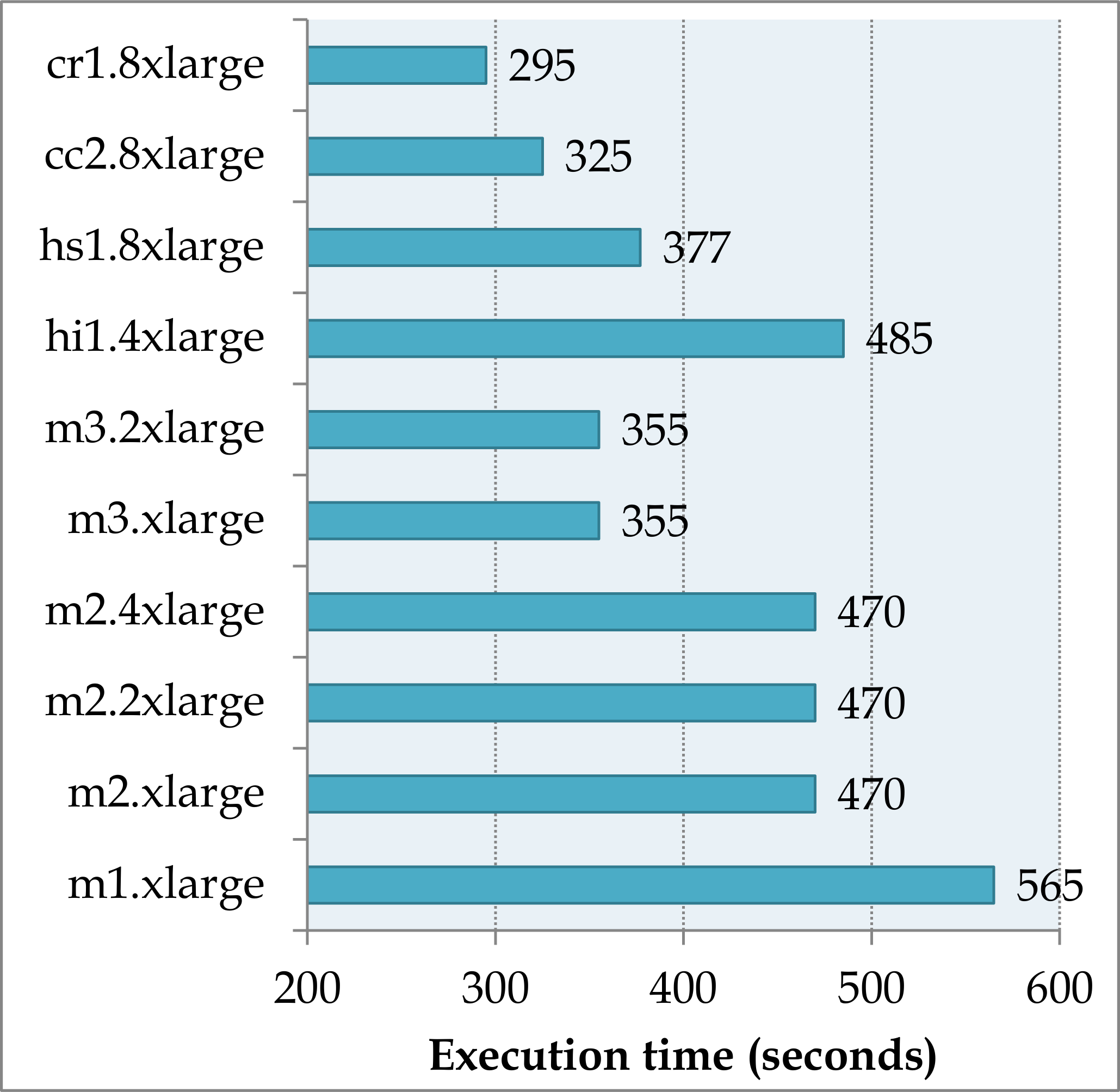}} \hfill
	\subfloat[Case study 3 - sequential]{\label{figure1-3}\includegraphics[width=0.325\textwidth]{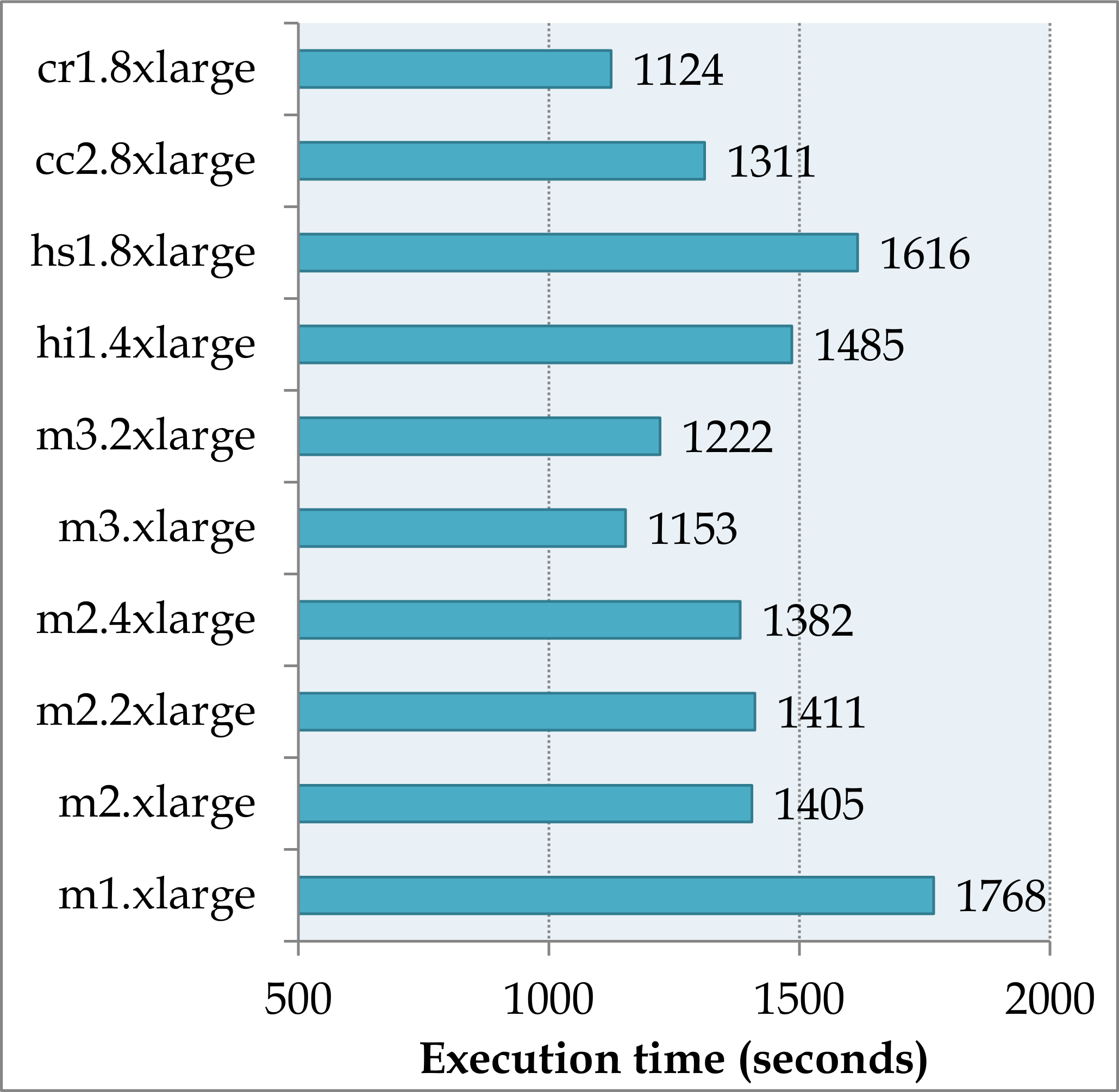}}\\
	\subfloat[Case study 1 - parallel]{\label{figure2-1}\includegraphics[width=0.325\textwidth]{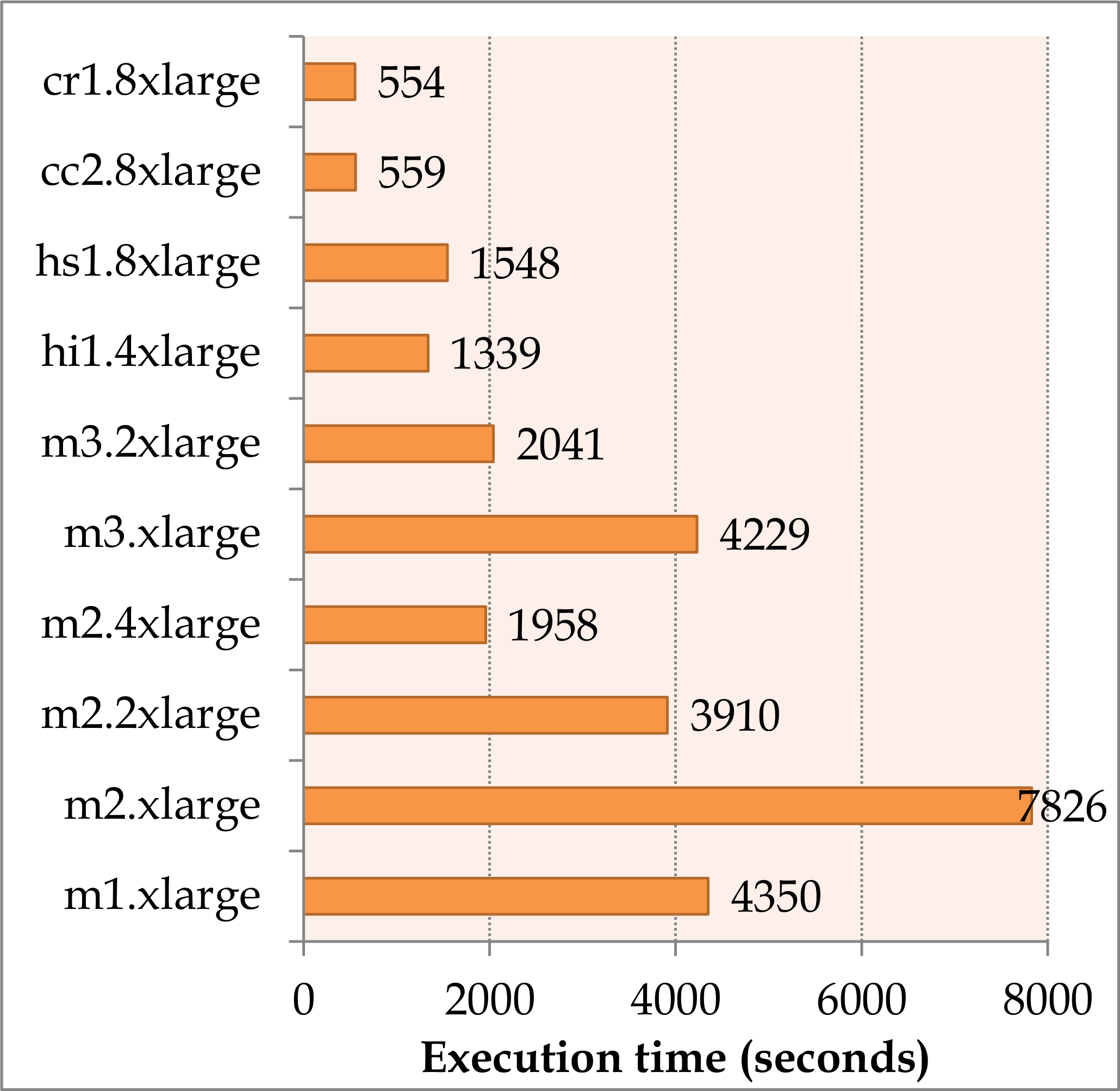}} \hfill
	\subfloat[Case study 2 - parallel]{\label{figure2-2}\includegraphics[width=0.325\textwidth]{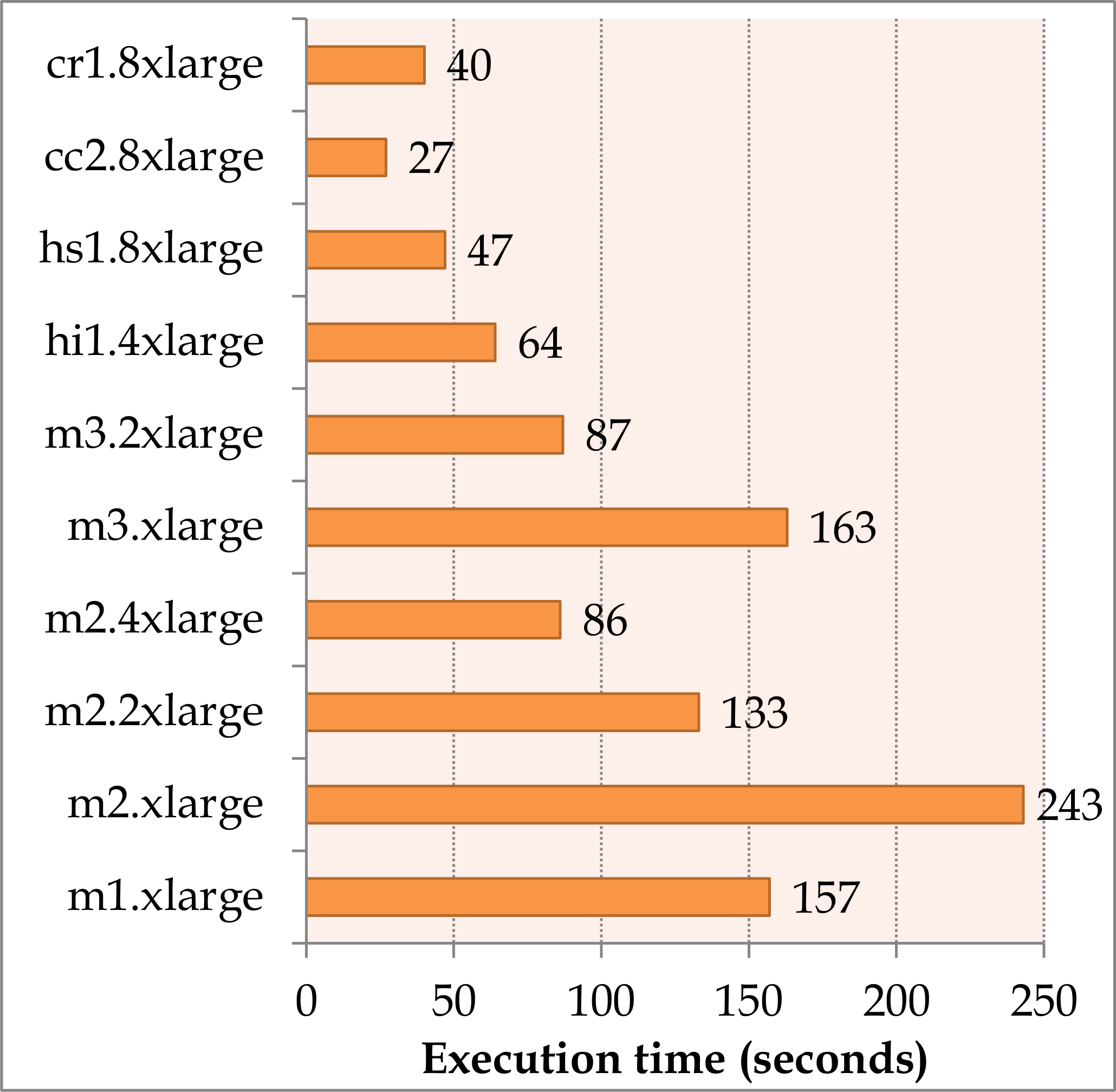}} \hfill
	\subfloat[Case study 3 - parallel]{\label{figure2-3}\includegraphics[width=0.325\textwidth]{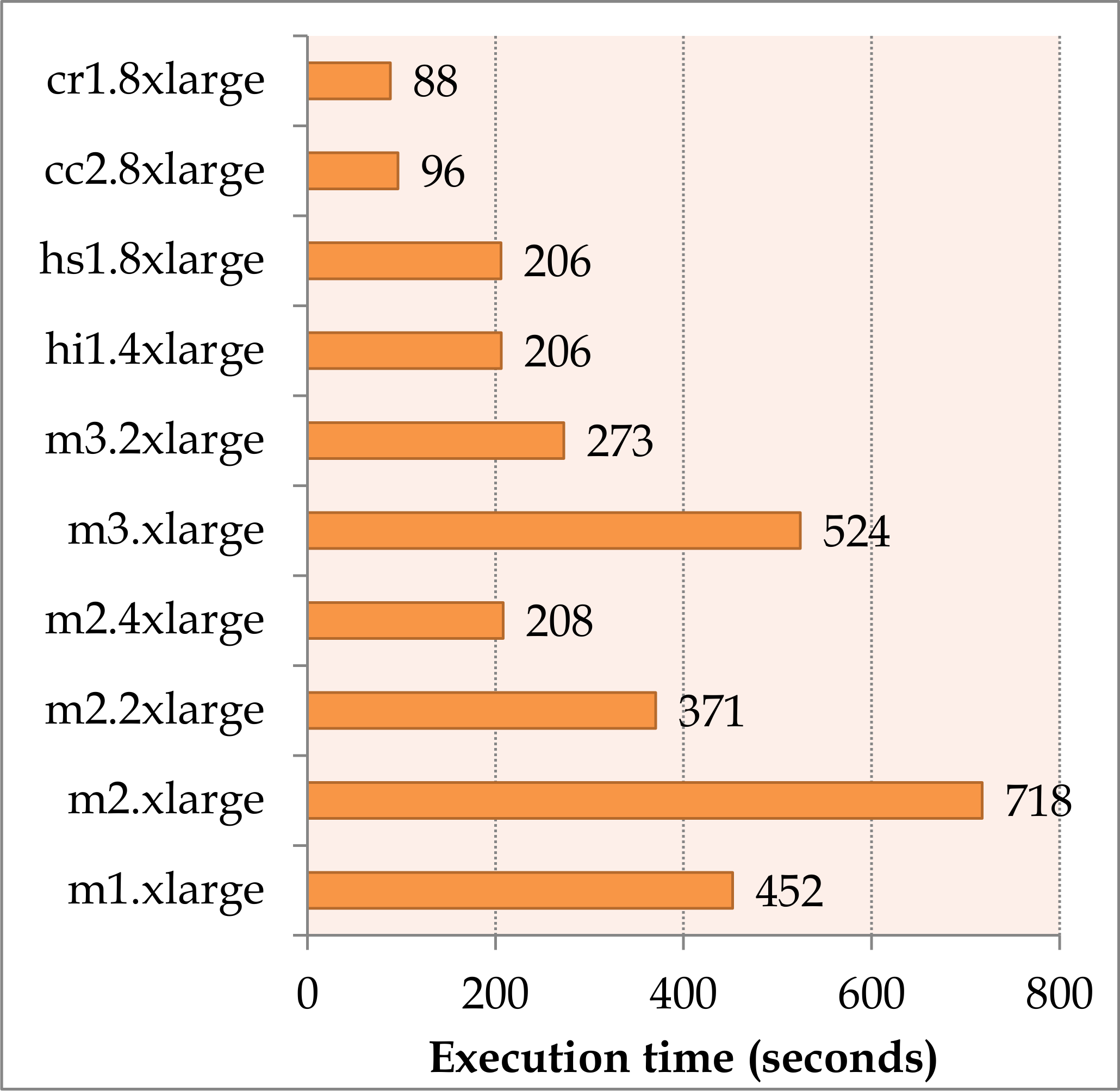}}
\caption{Sequential and parallel execution times for the case study applications}
\label{figure11}
\end{figure*}

In \textit{Step 1}, the three case study applications were executed on the VMs. The time taken to execute the application sequentially is presented in Figure \ref{figure1-1} to Figure \ref{figure1-3} and to execute the application in parallel using all available vCPUs is presented in Figure \ref{figure2-1} to Figure \ref{figure2-2}. In all case studies, the \texttt{cr1.8xlarge} and \texttt{cc2.8xlarge} have best performance; these VMs show good performance in memory and process and computation groups. The \texttt{m3} VMs are close competitors for sequential execution and \texttt{hi1.4xlarge} and \texttt{hs1.8xlarge} perform well for parallel execution. 
The results from parallel execution depend on the number of vCPUs available on the VM.

In \textit{Step 2}, the empirical ranks are generated using the standard competition ranking approach. The lowest time translates to the highest rank. If there are two VMs with the same program execution time they get the same rank and the ranking is continued with a gap. For example, in Figure \ref{figure1-1}, \texttt{m3.2xlarge} and \texttt{m3.xlarge} have the same program execution time. Both VMs have third rank and the next best performing VM, \texttt{hs1.8xlarge} obtains the fifth rank. 

In \textit{Step 3}, to generate the rankings from DocLite, a user provides the set of weights $W$ that characterise the case study applications, this is further explained in Section \cite{lightweight}. In consultation with domain scientists and practitioners, the weights for the three case studies are $\{4, 3, 5, 0\}$, $\{5, 3, 5, 0\}$ and $\{2, 0, 5, 0\}$ respectively. The above ranks were provided as input to the two benchmarking methods. 

Tables \ref{table2-1} to \ref{table2-3} show the empirical and benchmarking ranks for the three case studies using the lightweight container method as obtained in \textit{Step 4}. Tables \ref{table3-1} to \ref{table3-3} show the ranks for the case studies using the hybrid method (data from the lightweight container method along with data from the heavyweight method were considered). The historic benchmark data used in this paper was less than one month old when the hybrid method was executed.

Sequential and parallel ranks are generated for each case study using the weights. The empirical ranks are obtained from the timing results. The ranks obtained when using different sizes of the container are also reported in the tables. 

\begin{table}[h]
\centering
\begin{tabular}{ |p{1.4cm}|p{0.4cm}|p{0.4cm}|p{0.4cm}|p{0.4cm}|p{0.4cm}|p{0.4cm}|p{0.4cm}|p{0.4cm}|  }
\hline
\multirow{2}{*}{Amazon VM} & \multicolumn{4}{|c|}{Sequential Ranking} &
\multicolumn{4}{|c|}{Parallel Ranking}\\ \cline{2-9}
& Emp-irical & 100 MB & 500 MB & 1000 MB & Emp-irical & 100 MB & 500 MB & 1000 MB \\
\hline
\hline
m1.xlarge   & 9    &10     &10     &10     &9      &10     &10     &10 \\
m2.xlarge   & 7     &4      &4      &5      &10     &8      &8      &8 \\
m2.2xlarge  & 6     &7      &6      &7      &7      &9      &9      &9 \\
m2.4xlarge  & 5     &6      &7      &6      &5      &6      &6      &6 \\
m3.xlarge   & 4     &3      &3      &3      &8      &7      &7      &7 \\
m3.2xlarge  & 3     &5      &5      &5      &6      &4      &3      &4 \\
cr1.8xlarge & 1     &1      &1      &1      &1      &1      &1      &1 \\
cc2.8xlarge & 2     &2      &2      &2      &2      &2      &2      &2 \\
hi1.4xlarge & 8     &8      &8      &8      &3      &3      &4      &3 \\
hs1.8xlarge & 10     &9      &9      &9      &4      &5      &5      &5 \\
\hline
\end{tabular}
\caption{Case Study 1: Empirical and benchmark rankings for lightweight container benchmarking}
\label{table2-1}
\end{table}

\begin{table}[h]
\centering
\begin{tabular}{ |p{1.4cm}|p{0.4cm}|p{0.4cm}|p{0.4cm}|p{0.4cm}|p{0.4cm}|p{0.4cm}|p{0.4cm}|p{0.4cm}|  }
\hline
\multirow{2}{*}{Amazon VM} & \multicolumn{4}{|c|}{Sequential Ranking} &
\multicolumn{4}{|c|}{Parallel Ranking}\\ \cline{2-9}
& Emp-irical & 100 MB & 500 MB & 1000 MB & Emp-irical & 100 MB & 500 MB & 1000 MB \\
\hline
\hline
m1.xlarge 	&	10	&	10	&	10	&	10	&	8	&	10	&	10	&	10\\
m2.xlarge 	&	6	&	5	&	5	&	4	&	10	&	8	&	8	&	8\\
m2.2xlarge 	&	6	&	7	&	6	&	7	&	7	&	9	&	9	&	9\\
m2.4xlarge 	&	6	&	6	&	7	&	6	&	5	&	6	&	6	&	6\\
m3.xlarge 	&	3	&	3	&	3	&	3	&	9	&	7	&	7	&	7\\
m3.2xlarge 	&	3	&	4	&	4	&	5	&	6	&	4	&	4	&	4\\
cr1.8xlarge 	&	1	&	1	&	1	&	1	&	2	&	1	&	1	&	1\\
cc2.8xlarge 	&	2	&	2	&	2	&	2	&	1	&	2	&	2	&	2\\
hi1.4xlarge 	&	9	&	8	&	8	&	8	&	4	&	3	&	3	&	3\\
hs1.8xlarge 	&	5	&	9	&	9	&	9	&	3	&	5	&	5	&	5\\
\hline
\end{tabular}
\caption{Case Study 2: Empirical and benchmark rankings for lightweight container benchmarking method}
\label{table2-2}
\end{table}

\begin{table}[h]
\centering
\begin{tabular}{ |p{1.4cm}|p{0.4cm}|p{0.4cm}|p{0.4cm}|p{0.4cm}|p{0.4cm}|p{0.4cm}|p{0.4cm}|p{0.4cm}|  }
\hline
\multirow{2}{*}{Amazon VM} & \multicolumn{4}{|c|}{Sequential Ranking} &
\multicolumn{4}{|c|}{Parallel Ranking}\\ \cline{2-9}
& Emp-irical & 100 MB & 500 MB & 1000 MB & Emp-irical & 100 MB & 500 MB & 1000 MB \\
\hline
\hline
m1.xlarge 	&	10	&	10	&	10	&	10	&	8	&	10	&	10	&	10\\
m2.xlarge 	&	6	&	5	&	5	&	5	&	10	&	8	&	8	&	8\\
m2.2xlarge 	&	7	&	7	&	7	&	7	&	7	&	9	&	9	&	9\\
m2.4xlarge 	&	5	&	6	&	6	&	6	&	5	&	6	&	6	&	6\\
m3.xlarge 	&	2	&	3	&	3	&	3	&	9	&	7	&	7	&	7\\
m3.2xlarge 	&	3	&	4	&	4	&	4	&	6	&	5	&	5	&	5\\
cr1.8xlarge 	&	1	&	1	&	1	&	1	&	1	&	2	&	2	&	2\\
cc2.8xlarge 	&	4	&	2	&	2	&	2	&	2	&	1	&	1	&	1\\
hi1.4xlarge 	&	8	&	8	&	8	&	8	&	3	&	3	&	3	&	3\\
hs1.8xlarge 	&	9	&	9	&	9	&	9	&	3	&	4	&	4	&	4\\
\hline
\end{tabular}
\caption{Case Study 3: Empirical and benchmark rankings for lightweight container benchmarking}
\label{table2-3}
\end{table}

\begin{table}[h]
\centering
\begin{tabular}{ |p{1.4cm}|p{0.4cm}|p{0.4cm}|p{0.4cm}|p{0.4cm}|p{0.4cm}|p{0.4cm}|p{0.4cm}|p{0.4cm}|  }
\hline
\multirow{2}{*}{Amazon VM} & \multicolumn{4}{|c|}{Sequential Ranking} &
\multicolumn{4}{|c|}{Parallel Ranking}\\ \cline{2-9}
& Emp-irical & 100 MB & 500 MB & 1000 MB & Emp-irical & 100 MB & 500 MB & 1000 MB \\
\hline
\hline
m1.xlarge 	&	9	&	10	&	10	&	10	&	9	&	10	&	10	&	10\\
m2.xlarge 	&	7	&	5	&	5	&	5	&	10	&	9	&	9	&	9\\
m2.2xlarge 	&	6	&	7	&	7	&	7	&	7	&	8	&	8	&	8\\
m2.4xlarge 	&	5	&	6	&	6	&	6	&	5	&	6	&	6	&	6\\
m3.xlarge 	&	4	&	3	&	3	&	3	&	8	&	7	&	7	&	7\\
m3.2xlarge 	&	3	&	4	&	4	&	4	&	6	&	4	&	4	&	4\\
cr1.8xlarge 	&	1	&	1	&	1	&	1	&	1	&	1	&	1	&	1\\
cc2.8xlarge 	&	2	&	2	&	2	&	2	&	2	&	2	&	2	&	2\\
hi1.4xlarge 	&	8	&	8	&	8	&	8	&	3	&	3	&	3	&	3\\
hs1.8xlarge 	&	10	&	9	&	9	&	9	&	4	&	5	&	5	&	5\\
\hline
\end{tabular}
\caption{Case Study 1: Empirical and benchmark rankings for hybrid benchmarking}
\label{table3-1}
\end{table}

\begin{table}[h]
\centering
\begin{tabular}{ |p{1.4cm}|p{0.4cm}|p{0.4cm}|p{0.4cm}|p{0.4cm}|p{0.4cm}|p{0.4cm}|p{0.4cm}|p{0.4cm}|  }
\hline
\multirow{2}{*}{Amazon VM} & \multicolumn{4}{|c|}{Sequential Ranking} &
\multicolumn{4}{|c|}{Parallel Ranking}\\ \cline{2-9}
& Emp-irical & 100 MB & 500 MB & 1000 MB & Emp-irical & 100 MB & 500 MB & 1000 MB \\
\hline
\hline
m1.xlarge 	&	10	&	10	&	10	&	10	&	8	&	10	&	10	&	10\\
m2.xlarge 	&	6	&	5	&	5	&	5	&	10	&	9	&	9	&	9\\
m2.2xlarge 	&	6	&	7	&	7	&	7	&	7	&	8	&	8	&	8\\
m2.4xlarge 	&	6	&	6	&	6	&	6	&	5	&	6	&	6	&	6\\
m3.xlarge 	&	3	&	3	&	3	&	3	&	9	&	7	&	7	&	7\\
m3.2xlarge 	&	3	&	4	&	4	&	4	&	6	&	4	&	4	&	4\\
cr1.8xlarge 	&	1	&	1	&	1	&	1	&	2	&	1	&	1	&	1\\
cc2.8xlarge 	&	2	&	2	&	2	&	2	&	1	&	2	&	2	&	2\\
hi1.4xlarge 	&	9	&	8	&	8	&	8	&	4	&	3	&	3	&	3\\
hs1.8xlarge 	&	5	&	9	&	9	&	9	&	3	&	5	&	5	&	5\\
\hline
\end{tabular}
\caption{Case Study 2: Empirical and benchmark rankings for hybrid benchmarking}
\label{table3-2}
\end{table}

\begin{table}[h]
\centering
\begin{tabular}{ |p{1.4cm}|p{0.4cm}|p{0.4cm}|p{0.4cm}|p{0.4cm}|p{0.4cm}|p{0.4cm}|p{0.4cm}|p{0.4cm}|  }
\hline
\multirow{2}{*}{Amazon VM} & \multicolumn{4}{|c|}{Sequential Ranking} &
\multicolumn{4}{|c|}{Parallel Ranking}\\ \cline{2-9}
& Emp-irical & 100 MB & 500 MB & 1000 MB & Emp-irical & 100 MB & 500 MB & 1000 MB \\
\hline
\hline
m1.xlarge 	&	10	&	10	&	10	&	10	&	8	&	10	&	10	&	10\\
m2.xlarge 	&	6	&	5	&	5	&	5	&	10	&	9	&	9	&	9\\
m2.2xlarge 	&	7	&	7	&	7	&	7	&	7	&	8	&	8	&	8\\
m2.4xlarge 	&	5	&	6	&	6	&	6	&	5	&	6	&	6	&	6\\
m3.xlarge 	&	2	&	3	&	3	&	3	&	9	&	7	&	7	&	7\\
m3.2xlarge 	&	3	&	4	&	4	&	4	&	6	&	4	&	4	&	4\\
cr1.8xlarge 	&	1	&	1	&	1	&	1	&	1	&	1	&	1	&	1\\
cc2.8xlarge 	&	4	&	2	&	2	&	2	&	2	&	2	&	2	&	2\\
hi1.4xlarge 	&	8	&	8	&	8	&	8	&	3	&	3	&	3	&	3\\
hs1.8xlarge 	&	9	&	9	&	9	&	9	&	3	&	5	&	5	&	5\\
\hline
\end{tabular}
\caption{Case Study 3: Empirical and benchmark rankings for hybrid benchmarking}
\label{table3-3}
\end{table}

Given the rank tables for each case study it is important to determine the accuracy (or quality) of the ranks. In this paper, the accuracy of results is the correlation between the empirical ranks and the benchmark ranks. This quality measure validates the feasibility of using lightweight benchmarks and guarantees results obtained from benchmarking correspond to reality. 

\begin{table}[h]
\centering
\begin{tabular}{ |c|p{0.6cm}|p{0.6cm}|p{0.6cm}|p{0.6cm}|p{0.6cm}|p{0.6cm}|p{0.6cm}|p{0.6cm}|  }
\hline
\multirow{2}{*}{Case study} & \multicolumn{3}{|c|}{Sequential Ranking} &
\multicolumn{3}{|c|}{Parallel Ranking}\\ \cline{2-7}
& 100 MB & 500 MB & 1000 MB & 100 MB & 500 MB & 1000 MB \\
\hline
\hline
1 & 89.1 & 87.9 & 92.1 & 90.3 & 86.7 & 90.3\\
2 & 88.5 & 88.5 & 84.7 & 83.0 & 83.0 & 83.0\\
3 & 95.2 & 95.2 & 95.2 & 87.6 & 87.6 & 87.6\\
\hline
\end{tabular}
\caption{Correlation (in \%) between empirical and benchmarking ranks for the lightweight benchmarking method}
\label{table4-1}
\end{table}

\begin{table}[h]
\centering
\begin{tabular}{ |c|p{0.6cm}|p{0.6cm}|p{0.6cm}|p{0.6cm}|p{0.6cm}|p{0.6cm}|p{0.6cm}|p{0.6cm}|  }
\hline
\multirow{2}{*}{Case study} & \multicolumn{3}{|c|}{Sequential Ranking} &
\multicolumn{3}{|c|}{Parallel Ranking}\\ \cline{2-7}
& 100 MB & 500 MB & 1000 MB & 100 MB & 500 MB & 1000 MB \\
\hline
\hline
1 & 93.9 & 93.9 & 93.9 & 93.9 & 93.9 & 93.9\\
2 & 88.5 & 88.5 & 88.5 & 86.7 & 86.7 & 86.7\\
3 & 95.2 & 95.2 & 95.2 & 88.8 & 88.8 & 88.8\\
\hline
\end{tabular}
\caption{Correlation (in \%) between empirical and benchmarking ranks for the hybrid benchmarking method}
\label{table4-2}
\end{table}

In \textit{Step 5}, the correlation of the benchmark ranks using different containers and the empirical ranks for benchmarking is determined and shown in Table \ref{table4-1} and Table \ref{table4-2}; the percentage value shows the degree of correlation. Higher the correlation value the more robust is the benchmarking method since it corresponds more closely to the empirical ranks.

Consider Table \ref{table4-1}, on an average there is over 90\% and 86\% correlation between the empirical and benchmarked sequential and parallel ranks respectively. It is observed that increasing the size of the container does not generally increase the correlation between the ranks. The smallest container of 100 MB performs as well as the other containers. 

There is an average improvement of 1\%-2\% in the correlation between the ranks (Table \ref{table4-2}). While the hybrid method can improve the ranks, it is observed that the position of the top three ranks are not affected. Again, using the smallest container does not change the quality of results.   

\subsection{Summary} 
The experimental studies considered container benchmarking both in the context of varying the memory size and number of virtual cores of the VM. Variation of the number of cores is evaluated in the sequential (1 virtual core) and parallel (maximum number of virtual cores available) execution of the benchmarks. The results indicate that real-time benchmarking can be achieved which in turn will be useful for decision making for real-time deployments of applications on the cloud. This is substantiated by Figure \ref{evaluation:feasibility}; nearly 14 hours are required to benchmark a large VM entirely, however, using containers it can be done in just 8 minutes. This is significant improvement. 

The following three key observations are summarised from the experimental studies:

\begin{itemize}
\item[i.] Small containers using lightweight benchmarks perform similar to large containers. No improvement is observed in the quality of results with larger containers. On average, there is over 90\% and 86\% correlation when comparing ranks obtained from the empirical analysis and the 100 MB container.

\item[ii.] The hybrid method can slightly improve the quality of the benchmark rankings, although the position of the top three ranks do not change. The lightweight method is sufficient to maximise the performance of an application on the cloud. Implementing hybrid methods will require the storage of historic benchmark data and its maintenance over time.

\item[iii.] Since container-based benchmarks takes lower execution time compared to executing them directly on the VM they can be used for real-time deployment of applications.

\end{itemize}

\section{Conclusions}
\label{conclusions}
Benchmarking is important for selecting cloud resources that can maximise the performance of an application on the cloud. However, current benchmarking techniques are time consuming since they benchmark an entire VM for obtaining accurate benchmarks, which we have referred to as `\textit{heavyweight}', thereby limiting their real-time use for deploying an application. In this paper, we have explored an alternative to existing benchmarking techniques to generate accurate benchmarks in near real-time by using containers as a means to achieve `\textit{lightweight}' benchmarking. 

The fundamental assumption of this research is that light-weight benchmarking can produce comparable ranks to when an entire VM is benchmarked. This assumption is validated in this paper by generating ranks based on using containers and producing VM ranks which are compared against the ranks when an application is run on the entire VM. The correlation between the two sets of rank is found to be high which validates the assumption of this research. 

Docker Container-based Lightweight Benchmarking tool, referred to as `DocLite' was developed to facilitate lightweight benchmarking. DocLite organises the benchmark data into four groups, namely memory and process, local communication, computation and storage. A user of DocLite provides as input a set of four weights (ranging from 0 to 5), which indicate how important each of the groups are to scientific high-performance computing applications that needs to be deployed on the cloud. The weights are mapped onto the four benchmark groups and are used to generate a score for ranking the VMs according to performance. DocLite operates in two modes. In the first mode, containers are used to benchmark a portion of the VM to generate ranks of cloud VMs, and the second in which data obtained from the first mode is used in conjunction with historic data as a hybrid. DocLite is available to download from https://github.com/lawansubba/DoCLite. It is observed that benchmarking using DocLite is between 19-91 times faster than a heavyweight technique making them suitable for real-time. The experimental results highlight that the benchmarks obtained from container-based techniques are on an average over 90\% accurate making them reliable. Container-based technology is useful for benchmarking on the cloud and can form the basis for developing fast and reliable benchmarking techniques.

Container-based benchmarking techniques presented in this paper execute on Unix/Linux-based VMs regardless of whether they are provided by public/private clouds. We have tested these techniques on both previous and next generation VMs provided by Amazon although in this paper the results are presented for previous generation VMs. The techniques are repeatable by simply following the steps we have presented in Section \ref{lightweight} and Section \ref{hybrid}. However, the results obtained from the same type of VM may not be always the same, since performance of VMs are dependent on a number of factors including the workload at the data centre. 

We aim to extend our research in the future in the following four directions. Firstly, testing the benchmarking methodology on a wider range of applications as well as extending DocLite for monitoring the performance of the cloud VM executing an application to support dynamic scheduling.

Secondly, the benchmarking techniques in this paper focus on modelling the performance of a single VM using multiple cores. Hence, only memory and process, local communication, computation and storage are relevant without considering service level benchmarking that incorporates network benchmarks. Research that takes networking aspects into account has been explored previously \cite{net-1, net-2}. However, in this research we have not explored the integration of network benchmarks with the techniques proposed in this paper. By incorporating the networking aspect into the benchmarking the range of applications that can be benchmarked by the techniques can be widened.

Thirdly, the techniques we have proposed requires a user to provide a set of weights that describe the memory and process, computation, local communication and storage requirements of an application which need to be known beforehand. Hence, the benchmarking techniques are best suited for the class of problems, such as scientific high-performance computing workloads on the cloud, where the requirements of the application are known prior to deployment. The techniques can be significantly improved if the weights can be automatically determined using techniques such as profiling given the source code of the application that needs to be benchmarked on the cloud.

Fourthly, the hybrid method uses the same weight for both current and historic data. In this paper, the historic data is less than one month old; hence using the same weight generated good results as indicated by the correlation table. However, more historic data is likely to be stale and may not require the same weight as current benchmark data. We aim to consider the automatic generation of weights for historic data based on how stale it is. 

\section*{Acknowledgment}
This research was pursued under the EPSRC grant, EP/K015745/1, `Working Together: Constraint Programming and Cloud Computing,' an Erasmus Mundus Master's scholarship and an Amazon Web Services Education Research grant.

\balance



\end{document}